\documentclass[%
aip,
rsi,%
amsmath,amssymb,
reprint,%
]{revtex4-1}

\usepackage{graphicx}
\usepackage{dcolumn}
\usepackage{bm}
\usepackage{algorithm}%
\usepackage{algorithmicx}%
\usepackage{algpseudocode}%
\usepackage{float}
\usepackage[colorlinks=true, linkcolor=blue, citecolor=blue, urlcolor=blue]{hyperref}

\begin{document}
	
	\title{Power law scaling for classification accuracy in physical neural networks}

\author{Andrei V.Ermolaev}\email{andrei.ermolaev@femto-st.fr}
\affiliation{Universite Marie et Louis Pasteur, CNRS UMR 6174, Institut FEMTO-ST, 15B Avenue Montboucons, Besançon, 25000, France}

\author{Mathilde Hary}
\affiliation{Universite Marie et Louis Pasteur, CNRS UMR 6174, Institut FEMTO-ST, 15B Avenue Montboucons, Besançon, 25000, France}

\author{Anas Skalli}
\affiliation{Universite Marie et Louis Pasteur, CNRS UMR 6174, Institut FEMTO-ST, 15B Avenue Montboucons, Besançon, 25000, France}

\author{Go\"{e}ry Genty}
\affiliation{Photonics Laboratory, Tampere University, FI-33104 Tampere, Finland}

\author{Marcin Gebski}
\affiliation{Institute of Physics, Lodz University of Technology, ul. Wólczanska 219, 90-924 Lodz, Poland}

\author{Tomasz Czyszanowski}
\affiliation{Institute of Physics, Lodz University of Technology, ul. Wólczanska 219, 90-924 Lodz, Poland}

\author{Stephan Reitzenstein}
\affiliation{Technical University Berlin, Hardenbergstraße 36, D-10623 Berlin, Germany}

\author{James A. Lott}
\affiliation{Technical University Berlin, Hardenbergstraße 36, D-10623 Berlin, Germany}

\author{John M. Dudley}
\affiliation{Universite Marie et Louis Pasteur, CNRS UMR 6174, Institut FEMTO-ST, 15B Avenue Montboucons, Besançon, 25000, France}

\author{Daniel Brunner}
\affiliation{Universite Marie et Louis Pasteur, CNRS UMR 6174, Institut FEMTO-ST, 15B Avenue Montboucons, Besançon, 25000, France}

	\date{\today}
	
	\begin{abstract}
		
Physical neural networks (PNNs) harness the intrinsic complexity of physical systems to perform neural computation, potentially at speeds and energy efficiencies inaccessible to conventional digital hardware.
 Yet, a principled framework for quantifying and predicting their computing accuracy across diverse substrates has remained elusive.
 Here we introduce the Hotelling Trace Criterion (HTC), a task-dependent measure of PNN-state separability that can be evaluated without training.
 We demonstrate that it predicts PNN classification performance with high fidelity across highly nonlinear optical fibres, vertical-cavity surface-emitting lasers, and coupled nonlinear oscillator networks (CNON) , for benchmark tasks of different difficulty.
 Classification loss follows a power law in HTC, with Pearson correlation coefficients exceeding $0.99$ for MNIST and $\approx0.96$ for fashion MNIST.
 It is noteworthy that experimental and simulated data from physically distinct systems collapse onto a single scaling curve determined by the task rather than the substrate.
 Applying HTC layer-by-layer during training the CNON PNN further reveals that gradient-based optimisation distributes representational capacity unevenly across PNN layers, providing a quantitative diagnostic of training and architecture efficiency invisible to standard loss monitoring.
 Crucially, once the scaling exponent is established from a small number of trained calibration systems, all further performance predictions require no training since performance can be derived from the much more efficient HTC measurement.
 These results establish HTC as a substrate-agnostic figure of merit for comparing and scaling PNNs, advancing the field further towards a complete theory connecting fundamental hardware parameters to task performance through universal scaling laws.
		
	\end{abstract}
	
	\maketitle

\section*{Introduction}

Artificial neural networks (ANNs) are revolutionizing computing by tackling complex and abstract tasks once believed to be the exclusive domain of biological intelligence and human intellect.
 Yet, the relevance of ANN concepts goes far beyond applications only.
 A defining feature of ANNs is their ability to elevate data-driven optimization to unprecedented levels, rendering them highly adaptable for diverse applications, from sensory processing to edge computing.
 As a consequence, an ANN topology does not need to be perfectly pre-defined for a computational task, but it can be optimized in iterative steps using input data for which the correct result is know, called labelled data.
 This enabled implementing ANN inspired computing in non-digital, complex physical systems which offer superior performance in energy efficiency, speed, parallelism and latency, for example \cite{markovic2020physics}.
 
Such computing with physical systems inspired by ANN principles, i.e. physical neural networks (PNNs) has a history stretching back nearly four decades, with optics as the pioneering substrate~\cite{psaltis1985}.
 In this context, a particularly productive PNN framework has been reservoir computing \cite{appeltant2011information}, in which only a PNN's output weights are trained, while the physical system's topology itself remains fixed, making dynamics of nonlinear physical systems immediately exploitable as a computing resource.
 This paradigm has since been realized across domains, including optoelectronic~\cite{larger2012}, all-optical~\cite{brunner2013parallel}, spintronic~\cite{torrejon2017,romera2018}, and mechanical systems~\cite{dion2018}, each potentially offering distinctive advantages.
 The field continuous to advance rapidly, and a wide range of PNNs without reservoir computing topology restrictions can now be \emph{programmed} using data-driven optimization of their topology \cite{wright2022deep,skalli2025,momeni2025training} to execute complex computations with unparalleled key performance metrics \cite{shastri2021photonics}.
 
However, current PNNs remain too small and their topology too simple to tackle technologically and societally relevant tasks for which digital alternatives are ill-suited.
 Advancing PNN computing towards larger systems with advanced topologies requires significant future investment of time and other resources \cite{skalli2026thin}.
 In digital computing, such investments became justifiable when predictable scaling for the field's development was established through Moore’s Law, while in artificial intelligence (AI), the discovery of power law scaling between compute and performance for large language models ~\cite{kaplan2020scaling,hoffmann2022training} served as a similar catalyst.
 In this context scaling laws are therefore significant for two reasons: (i) they shed light on potential underlying mechanisms of a system, and (ii) they strengthen investment confidence among both government and private actors.
 Accelerating the relevance and development of PNNs therefore critically depends on identifying comparable scaling laws.

Uncovering these scaling laws for PNNs poses a significantly greater challenge as compared to classical digital computing architectures.
 Physical neural networks typically leverage continuous state-space dynamics for computation, unlike digital computing, where countable computational units such as floating point operations (FLOPs) for digital computing and multiply and accumulate (MACs) for ANN computing are directly defined by hardware design.
 In PNNs computation is not performed through discrete, clearly separated digital operations, but through complex and typically continuos-valued responses of a physical system.
 Additionally, different physical degrees of freedom may contribute to the computation in different ways, which makes it difficult to quantify the computational capacity of such systems.
 As a result, developing a reliable predictor of PNN performance for specific tasks has proven difficult and is so far lacking.
 Previous approaches have taken initial steps in both theory \cite{dambre2012information,tureci2023} and experiments \cite{Skalli2022,hary2025}, but they were constrained to single systems and did not establish a general link between a computational capacity metric and task-specific PNN computing performance.
 
In this work, using experimental and numerical data from several PNN systems, we demonstrate precisely such a metric for classification tasks.
 We analyse a PNN’s state-space dynamics in response to input information using the Hotelling Trace Criterion (HTC), an algorithm that normalizes distances between groups of data-points by the respective size of their distribution~\cite{Hotelling1951AGT}.
 This provides a class-specific separability metric, and we demonstrate its power law scaling relationship with PNN classification accuracy using both experimental and numerically simulated PNNs.
 Experimental data include PNNs implemented via nonlinear propagation of ultrashort optical pulses in optical fibres \cite{hary2025} and nonlinear transformations obtained from a high-dimensional semiconductor multimode laser \cite{skalli2025}.
 We complement these experiments with numerical simulations of the ultrashort optical pulse PNN, as well as with a multi-layer coupled nonlinear oscillator network (CNON), serving as a generic, substrate-independent PNN model system.
 PNN computing performance is evaluated using the MNIST and fashion MNIST benchmark classification tasks, and notably, for each task, all PNN systems collapse onto essentially the same scaling law.

Our findings have far-reaching consequences.
 On a fundamental level, they establish a direct link between physical systems and their task-specific computational capacity, thereby laying the groundwork for new and additional PNN computing metrics.
 Unlike rank, kernel quality, and memory capacity metrics, which have limited relevance for real-world tasks due to their unclear connection to task-specific performance, our approach is inherently task-focused.
 Furthermore, this work opens new avenues for the quantitative evaluation of training methods and their potential future improvement.
 Finally, these findings contribute to the development of a general theory for PNN computing~\cite{jaeger2023toward} by establishing the first link between PNN classification capacity and performance.

\section*{Results}

\subsection*{Physical Neural Network computing}
\label{sec_sub:PNN_comp}

\begin{figure}[ht]
\begin{center}
\includegraphics[width=0.8\linewidth]{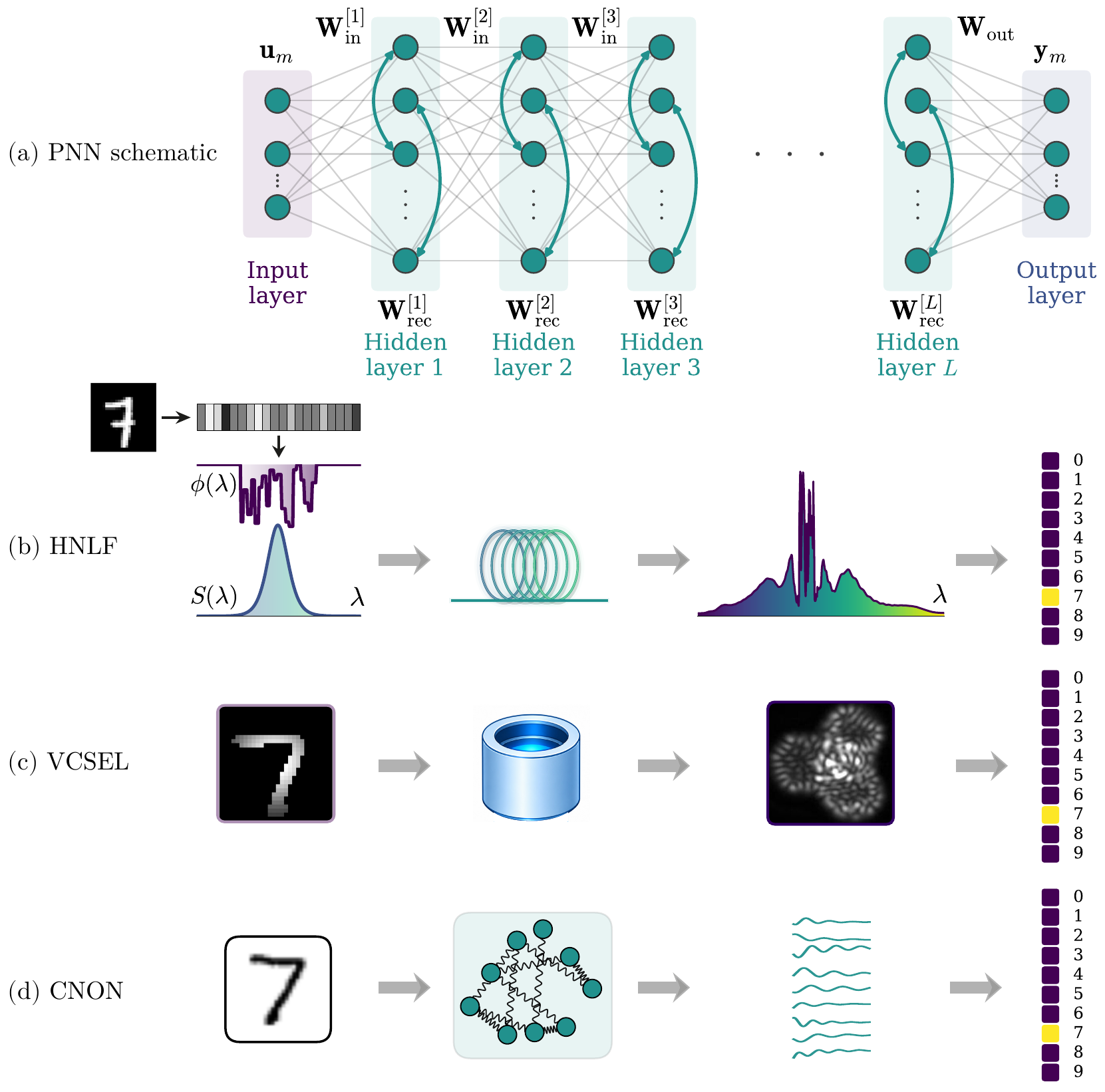}
\caption{\textbf{Schematic illustration of PNN platforms considered in this work.}
\textbf{(a)} Generic multilayer artificial neural network (ANN) standing in as structural inspiration of physical neural networks.
\textbf{(b)} Highly-nonlinear optical fibre (HNLF)-based PNN. A spectral phase mask $\phi(\lambda)$ applied to input spectral amplitude $S(\lambda)$ creates input information. Nonlinear propagation through the HNLF transforms this into a broad and high-dimensional output spectrum that constitutes the hidden-layer representation.
\textbf{(c)} Large area vertical-cavity surface-emitting laser (LA-VCSEL) based PNN. A narrow-linewidth laser illuminates a spatial light modulator displaying the input data which is then injected into the LA-VCSEL which creates the hidden-layer response.
\textbf{(d)} The coupled nonlinear oscillators network (CNON) is the PNN equivalent of the ANN in \textbf{(a)}. The steady-state temporal responses of individual oscillators act as activation functions.
}
\label{fig:PNNs}
\end{center}
\end{figure}

The continuing success of ANNs exposed a deepening mismatch between algorithmic ambition and hardware capability.
 Classical digital electronic processors struggle to meet the computational demands of state-of-the-art ANN models, and PNNs aim to address this challenge at its root by embedding the essential elements of neural computation: nonlinearity, high dimensionality, and connectivity directly into the intrinsic dynamics of a physical system, as illustrated in Fig.~\ref{fig:PNNs}(a).

 An external signal injected into the physical system constitutes the PNN's input $\mathbf{u}_m$, where $m = 1, \dots, M$ indexes individual input examples.
 The system's response across its many degrees of freedom then plays the role of ANN hidden-layer activations, with each state-space analogous to an ANN neuron.
 These activations arise from direct coupling to the external drive signal and internal interactions between physical dimensions, which are analogues of ANN synaptic connections.
 A compact description of PNN hidden-layer dynamics and readout can, for example, be given by the ordinary differential equations
%
%
\begin{align}
    \tau \frac{d\mathbf{h}_m}{dt} &= -\mathbf{h}_m + f\!\left(\mathbf{W}_{\textrm{rec}}\,\mathbf{h}_m + \mathbf{W}_{\mathrm{in}}\,\mathbf{u}_m\right),
    \label{eq:PNN_ODE} \\
    \mathbf{y}_m &= \mathbf{W}_{\mathrm{out}}\,\mathbf{h}_m,
    \label{eq:PNN_readout} \\
    \mathcal{L} &= \frac{1}{M}\sum_{m=1}^{M}\left\|\mathbf{y}_m - \mathbf{y}_m^{\mathrm{target}}\right\|^2,
    \label{eq:MSE}
\end{align}
\noindent where $\tau$ is the characteristic relaxation timescale of the physical system, $f(\cdot)$ is the activation function emerging from the PNN's intrinsic nonlinearity, $\mathbf{W}_{\textrm{rec}}$ is the internal connectivity matrix, $\mathbf{W}_{\mathrm{in}}$ couples the external input into the hidden-layer state $\mathbf{h}_m$, and $\mathbf{W}_{\mathrm{out}}$ projects the hidden-layer state to the output $\mathbf{y}_m$.
 In the context of PNNs, during the injection of one input data sample input information $\mathbf{u}_{m}$ remains constant, while hidden layer state $\textbf{h}_{m}$ evolves freely according to Eq.~\ref{eq:PNN_ODE}, and the system performance is measured by a loss functions, here the mean square error (MSE) loss $\mathcal{L}$ given in Eq.~{\ref{eq:MSE}.
 In direct analogy to ANN training, optimizing a PNN for a specific task requires that at least some aspects of the topology encoded in $\mathbf{W}_{\textrm{rec}}$, $\mathbf{W}_{\mathrm{in}}$ and $\mathbf{W}_{\mathrm{out}}$ are tunable in order to improve performance by reducing $\mathcal{L}$.
 For the MNIST handwritten digit classification benchmark, for example, this amounts to adjusting the system's internal structure such that injection of any image of a given digit class causes a designated dimension of $\mathbf{y}_m$ to attain the highest activation value.

Our first system exploits the broadband nonlinear dynamics arising during femtosecond pulse propagation in a highly-nonlinear optical fiber (HNLF), as reported in Ref.~\cite{Ermolaev-2025,hary2025} and illustrated in Fig.~\ref{fig:PNNs}(b).
 An input is encoded by applying a spectral phase mask $\phi(\omega)$ to the spectral amplitude $S(\omega)$. The corresponding pulse $A_0(T) = \mathcal{F}^{-1}\{S(\omega - \omega_0) \exp[i \phi(\omega - \omega_0)]\}$ is then launched into the HNLF, where the dynamics is governed by the complex interplay of dispersive and nonlinear effects governed by the generalised nonlinear Schr\"{o}dinger equation (GNLSE)
\begin{equation}
    i \frac{\partial A}{\partial z} +\sum_{j \geq 2} \dfrac{i^{j}}{j!}\beta_{j} \dfrac{\partial^j{A}}{\partial{T}^j} + \gamma \left( 1 + \dfrac{i}{\omega_0} \, \frac{\partial}{\partial T} \right) \left( A\,[R \ast |A|^2 ]\right) = 0, 
    \label{eq:GNLSE_main}
\end{equation}
\noindent where $A(z,T)$ is the complex field envelope as a function of distance $z$ and co-moving time $T$, $\beta_j$ stands for dispersion coefficients, $\gamma$ is the nonlinear coefficient, and $\omega_0$ is the carrier frequency.
 The nonlinear response function in the convolution term ($\ast$) represents the instantaneous Kerr and delayed Raman response of fused silica~\cite{Dudley-2006} (see Method section for details).
 The spectral output, sampled across wavelength bins corresponds to the PNN's hidden state vector $\mathbf{h}_m$~\cite{hary2025}.
 This system is particularly well-suited to numerical reproduction via direct GNLSE simulation, and we leverage this to complement and interpret experimental results~\cite{Ermolaev-2025}.

Our second system is a large-area vertical-cavity surface-emitting laser (LA-VCSEL)~\cite{haghighi202040} that was leveraged as PNN~\cite{skalli2025} illustrated in Fig.~\ref{fig:PNNs}(c).
 Input images are encoded onto an optical injection field that drives the LA-VCSEL, whose spatially distributed transverse lasing modes undergo complex nonlinear modal interactions and form the network's hidden layer.
 The resulting multimode emission pattern is multiplied by a trained readout layer $\mathbf{W}_{\mathrm{out}}$ to obtain the classification result.
 Crucially, the rich spatiotemporal laser dynamics involve coupled electromagnetic fields, carrier dynamics, and thermal effects across potentially thousands of spatial modes, rendering a tractable first-principles numerical model inaccessible.
 This very intractability underscores one of the core PNN arguments: the LA-VCSEL autonomously executes, in a single optical transit, a transformation whose digital simulation would demand computational resources far exceeding those of the PNN physical implementation, directly embodying the efficiency argument for PNN hardware.

Our third system is the numerical CNON toy model, as illustrated in Fig.~\ref{fig:PNNs}(d).
 Each node in the network is a nonlinear oscillator coupled to its neighbours and to the external input, with the network state evolving according to the general PNN dynamics of Eq.~\eqref{eq:PNN_ODE} (see Sec.~\ref{sec:CNON} for details).
 This system makes no claim to a particular physical substrate, but instead its role is to stand in as interpretable  performance reference across PNN implementations.
 Unlike the HNLF or LA-VCSEL systems, it enables the study of PNN computing for systems across a wide range of network sizes, coupling topologies as well as nonlinearity parameters.
 Such variability is often not attainable for PNNs based on experiments or physically correct numerical simulations, as those are typically constraint in size as well as available parameter ranges.
 Like Kuramoto oscillator networks for general observations of synchronization and nonlinear dynamics in physical systems, our CNON PNN therefore serves as a generic and analytically transparent representative of PNN dynamics and its connection to PNN computing performance and metrics.

\section*{Computational performance predictors}

Several complementary approaches to quantifying a PNN's maximum dimensionality have been proposed, each leveraging the PNN's collected responses towards a large sample of input data examples.
 Dambre~et~al.\ introduced the information processing capacity framework, in which the total computational capacity of a dynamical system within the framework of PNN regression~\cite{appeltant2011information,Manuylovich2025} is quantified by expanding its input-output map onto an orthonormal basis of Legendre polynomials~\cite{dambre2012information}.
 A complementary framework was introduced by Hu~et~al.\, who defined the resolvable expressive capacity of a physical system under finite measurement noise~\cite{tureci2023} by extracting a basis of eigentasks, quantifying the orthogonal input-output functions a PNN can resolve above its noise floor.
 Finally, PNN dimensionality analysis based on principal component analysis (PCA) performs singular value spectrum computations of the PNN's response to a sequence of random inputs, quantifying how many statistically independent dimensions this activates inside the system.
 The approach was applied to the HNLF~\cite{hary2025} and LA-VCSEL~\cite{skalli2022computational} systems that are also studied in our current work.
 Similar to the eigentask approach, PCA analysis was further generalized by introducing an indicator function identifying the number of PCA dimensions exceeding the system's noise-floor~\cite{skalli2022computational}, which recently was extended to non-uniform noise~\cite{ramachandran2026information}.

The central challenge we aim to address in our manuscript is how to generically characterize PNNs and their number of  computationally relevant degrees of freedom within the context of a particular computational task.
 A PNN whose dynamics collapse onto a low-dimensional state, regardless of how many physical degrees of freedom it nominally possesses, offers no more classification capacity than a much smaller ANN.
 Contrary to that, a high-dimensional response could potentially enable the PNN be able to solve complex classification tasks.
 However, these purely dimensional-based metrics such as the ones introduced before fail to capture if dynamics induced by inputs belonging to different input data classes strongly overlap in their response along those dimensions.
 In such a case, PNN dimensionality will be high yet still result in poor classification accuracy reflected in a high loss function value $\mathcal{L}$.
 The consequence of this situation can be observed in \cite{hary2025}, where an increase in PCA dimensionality did not result in improved classification accuracy.

\subsection*{Hotelling trace criterion as the separability metric}
\label{sec:HTC}

Addressing the aforementioned shortcoming in dimensionality-only metrics requires expanding those in order to include a measure quantifying the overlap of PNN responses belonging to different classes.
 Here, we refer to the Hotelling trace criterion (HTC), which is a statistical measure of class separability first introduced in the context of multivariate generalizations of Student's $t$ and Fisher's $z$-tests~\cite{Hotelling1931,Lawley1938}.
 Besides many applications in statistical analysis~\cite{Fukunaga1972IntroductionTS,Bishop2008}, HTC was shown to be an effective scalar metric for optimizing imaging systems and to strongly correlate with human-observer performance~\cite{Smith1986,Fiete1987}.
 In machine learning, it is usually seen as the separability criterion maximized by linear discriminant analysis~\cite{Rao1963,Tharwat2017}, and it has more recently been used to improve class separation in the contexts of feature selection, image recognition, discriminative clustering, and multilayer ELMs classifiers~\cite{Gu2012,Iosifidis2013,Huang2015,Wu2017,Lai2022}.

HTC can be directly computed from the hidden layer matrix $\mathbf{H}$.
 Let $\mathbf{H} \in \mathbb{R}^{M \times N}$, where $M$ is the number of samples (images) and $N$ is the dimensionality or number of neurons.
 Then, each row of $\mathbf{H}$ corresponds to a network state vector of a hidden layer $\mathbf{h}_m^{T}$, for $m = 1, \dots, M$. 
 For $K$ classes with $M_k$ samples in each, the class mean hidden layer representation is defined as
\begin{equation}
\boldsymbol{\mu}_k = \frac{1}{M_k} \sum_{m \in k}^{K} \mathbf{h}_m ,
\label{eq:k-mean}
\end{equation}
while the global mean representation across all samples is
\begin{equation}
\boldsymbol{\mu} = \frac{1}{M} \sum_{m=1}^{M} \mathbf{h}_m = \frac{1}{M}\sum_{k=1}^{K} M_k\boldsymbol{\mu}_k.
\label{eq:g-mean}
\end{equation}

Next, in analogy to multiclass linear discriminant analysis we define the between-class scatter matrix that measures the variance of the class means Eq.~(\ref{eq:k-mean}) relative to the global mean Eq.~(\ref{eq:g-mean})
\begin{equation}
\mathbf{S}_B =
\sum_{k=1}^{K}
M_k (\boldsymbol{\mu}_k - \boldsymbol{\mu})(\boldsymbol{\mu}_k - \boldsymbol{\mu})^T ,
\end{equation}
and the within-class scatter matrix that quantifies the cumulative variance of samples around their respective class means
\begin{equation}
\mathbf{S}_W =
\sum_{k=1}^{K} \sum_{m \in k}
(\mathbf{h}_m - \boldsymbol{\mu}_k)(\mathbf{h}_m - \boldsymbol{\mu}_k)^T .
\end{equation}
Finally, the HTC is defined as a scalar separability metric by normalizing the between-class scattering matrix by the within-class scatter matrix:
\begin{equation}
\mathrm{HTC} = \mathrm{tr}\left(\mathbf{S}_W^{-1}\mathbf{S}_B\right) = \sum_{i}\lambda_{i}.
\label{eq:HTC}
\end{equation}
Here, $\mathrm{tr}$ denotes the trace of a matrix.
 Note that the HTC is equivalent to the sum of non-zero eigenvalues $\lambda_{i}$, representing the ratio of between-class to within-class variance along principal directions, and consequently captures the geometric aspects for achieving a higher probability that the examples within one class are correctly divided by the linear separatrix implemented through $\mathbf{W}_{\mathrm{out}}$.
 An increasing $\mathrm{HTC}$ indicates an increasing separation between class centroids relative to the within-class variance.

\begin{figure}[H]
\begin{center}
\includegraphics[width=0.8\linewidth]{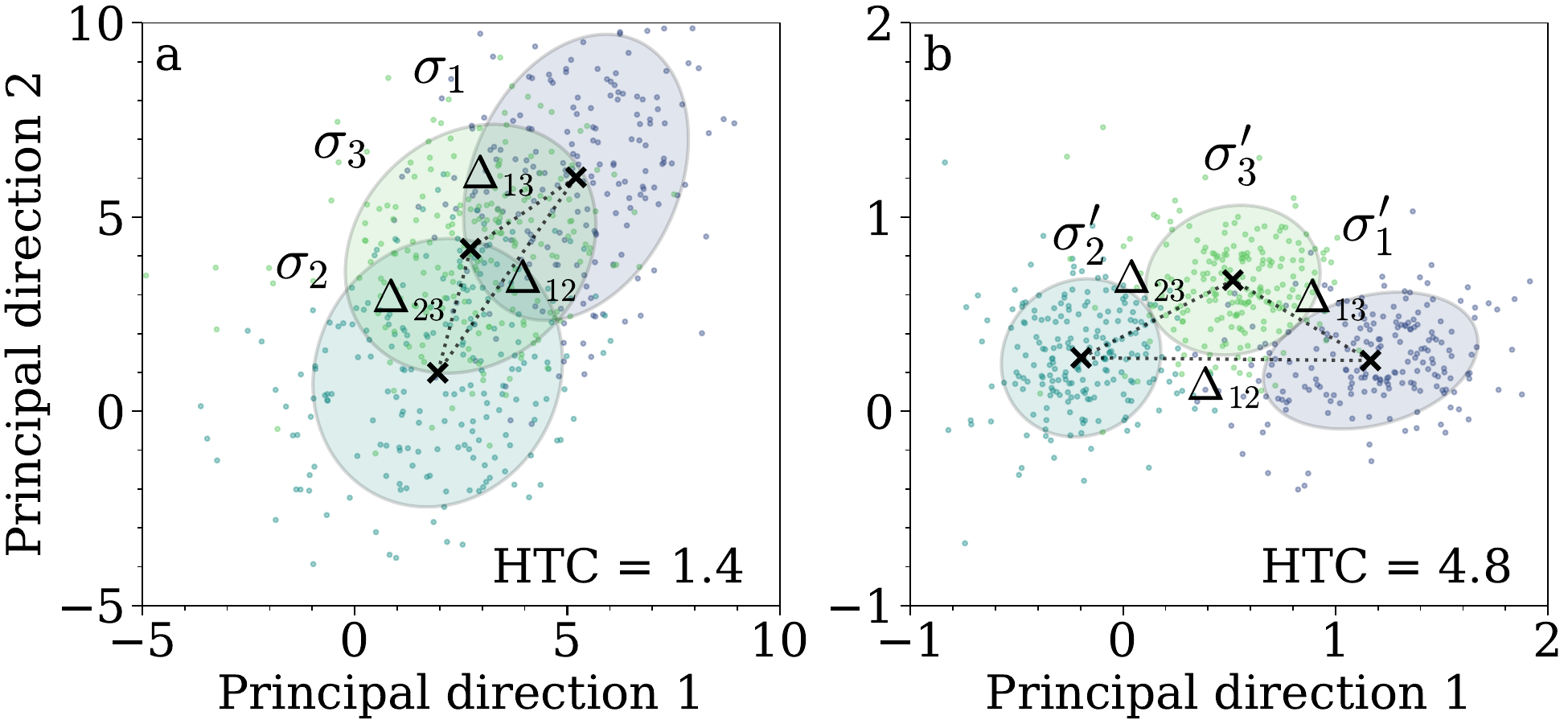}
\caption{\textbf{Illustration of the Hotelling trace criterion metric used to quantify the class separability.} \textbf{(a, b)} Projection of hidden-layer representations of the CNON onto two principal directions of $\mathbf{S}_W^{-1}\mathbf{S}_B$, shown for three representative MNIST classes in a low-performance ($\alpha=0.15$) and high-performance ($\alpha=0.5$) configurations, respectively. Crosses mark class centroids, $\Delta_{ij}$ represent the distances between cluster centroids, and $\sigma_i$ and $\sigma'_i$ denote the total within-class variance of each class cluster. Despite larger distances between centroid in \textbf{(a)}, the substantially larger total variances yield a lower $\mathrm{HTC}$ value (1.4 versus 4.8).}
\label{fig:Projection}
\end{center}
\end{figure}

In Fig.~\ref{fig:Projection}\textbf{(a, b)} we show the projection of hidden-layer representations onto the two first principal directions of $\mathbf{S}_W^{-1}\mathbf{S}_B$ for a low-performance \textbf{(a)} and a high-performance \textbf{(b)} PNN, with data obtained from the numerical CNON model, Fig.~\ref{fig:PNNs}\textbf{(d)}, for weaker ($\alpha=0.15$) and stronger damping ($\alpha=0.5$), respectively.
 For clarity we restricted our HTC projection to three classes in the input-data corresponding to MNIST handwritten digits 0, 3, 6.
 In order to provide an intuitive explanation, we leverage that for the simplification of isotropic (spherical) and balanced classes (same number of samples) projected to $N$-dimensional space, Eq.~(\ref{eq:HTC}) simplifies to
\begin{equation}
  \mathrm{HTC} = \operatorname{tr}\!\left(S_W^{-1} S_B\right)
  = \dfrac{N}{K}\,\dfrac{\sum_{j<k}\Delta_{jk}^{2}}{\sum_{k=1}^{K}\sigma_k^{2}},
  \label{eq:htc_balanced}
\end{equation}
where $\sigma_k = (\sum_{n=1}^{N}\sigma_{kn}^{2})^{1/2}$ is the total within-class
variance of class $k$ with $\sigma_{kn}$ denoting the variance of class $k$ in the direction of $n$, $\Delta_{jk} = \lVert \boldsymbol{\mu}_j - \boldsymbol{\mu}_k \rVert$ is the
Euclidean distance between the centroids of classes $j$ and $k$, and
$\sum_{j<k}$ denotes the sum over all distinct pairs of class centroids. For the three classes projected onto two principal directions ($K=3$, $N=2$), shown in Fig.~\ref{fig:Projection}\textbf{(a,b)}, this further simplifies to
\begin{equation}
  \mathrm{HTC} = \frac{2}{3}\,
  \frac{\Delta_{12}^{2} + \Delta_{13}^{2} + \Delta_{23}^{2}}
       {\sigma_{1}^{2} + \sigma_{2}^{2} + \sigma_{3}^{2}},
  \label{eq:htc_3class}
\end{equation}

\noindent which clearly shows the underlying idea of the HTC metric, which accounts for the ratio of total between-class separation and within-class variance. Indeed, while class centroids are spaced far apart in Fig.~\ref{fig:Projection}\textbf{(a)} and, hence, yield higher between-class distances, simultaneously the three class clusters exhibit substantial overlap due to large total variances (denoted $\sigma_j$), yielding a low $\mathrm{HTC}=1.4$ in the projected two-dimensional subspace associated with a MSE loss of $\mathcal{L}=0.055$.
 In Fig.~\ref{fig:Projection}\textbf{(b)}, closer spaced clusters are however combined with an even smaller variance and overlap (denoted $\sigma_j'$), resulting in clearer class separation reflected in a higher $\mathrm{HTC}=4.8$ and achieving a $\mathcal{L}=0.035$.
 Equation~(\ref{eq:htc_3class}) simplifies the full HTC metric by assuming isotropic clusters, i.e. spherical instead of elliptical clusters.
 These simplifications serve an illustrative purposes linked to Fig.~\ref{fig:Projection}, and for the rest of the paper we use Eq.~(\ref{eq:HTC}) to compute HTC.
 The conceptual link between HTC and the data-wise computed loss $\mathcal{L}$ is the following.
 HTC characterises the PNN response clusters induced by different input classes by approximating them as high-dimensional ellipsoids within the orthogonal state-space, and then quantifies their separability by normalising inter-cluster distances by the respective cluster widths before summing over all class pairs.
 The classification loss, by contrast, computes a precise point-by-point distance in the solution space after projecting the PNN's response through the linear output layer $\mathbf{W}_{\textrm{out}}$.
 
\subsection*{Performance comparability between different PNNs}

We now link this metric to the classification performance of mulitple PNNs.
 Figures~\ref{fig:HTC_PNNs}\textbf{(a, b, c)} show the test and training accuracies for all ten classes of the MNIST handwritten digit task obtained for the three considered PNNs as functions of their physical parameters.
 Throughout the manuscript, data originating from experiments (numerical simulations) are presented using filled (open) symbols.
 The results in Fig.~\ref{fig:HTC_PNNs} treat each PNN in the reservoir computing and ELM frameworks, respectively, meaning that only the readout matrix $\mathbf{W}_{\mathrm{out}}$ is optimized during offline training (further details are provided in the Methods section).
 All three systems exhibit well-defined performance optima at specific physical parameter values, showing that PNN computing capacity is strongly governed by the physical operating point.

\begin{figure}[H]
\begin{center}
\includegraphics[width=0.9\linewidth]{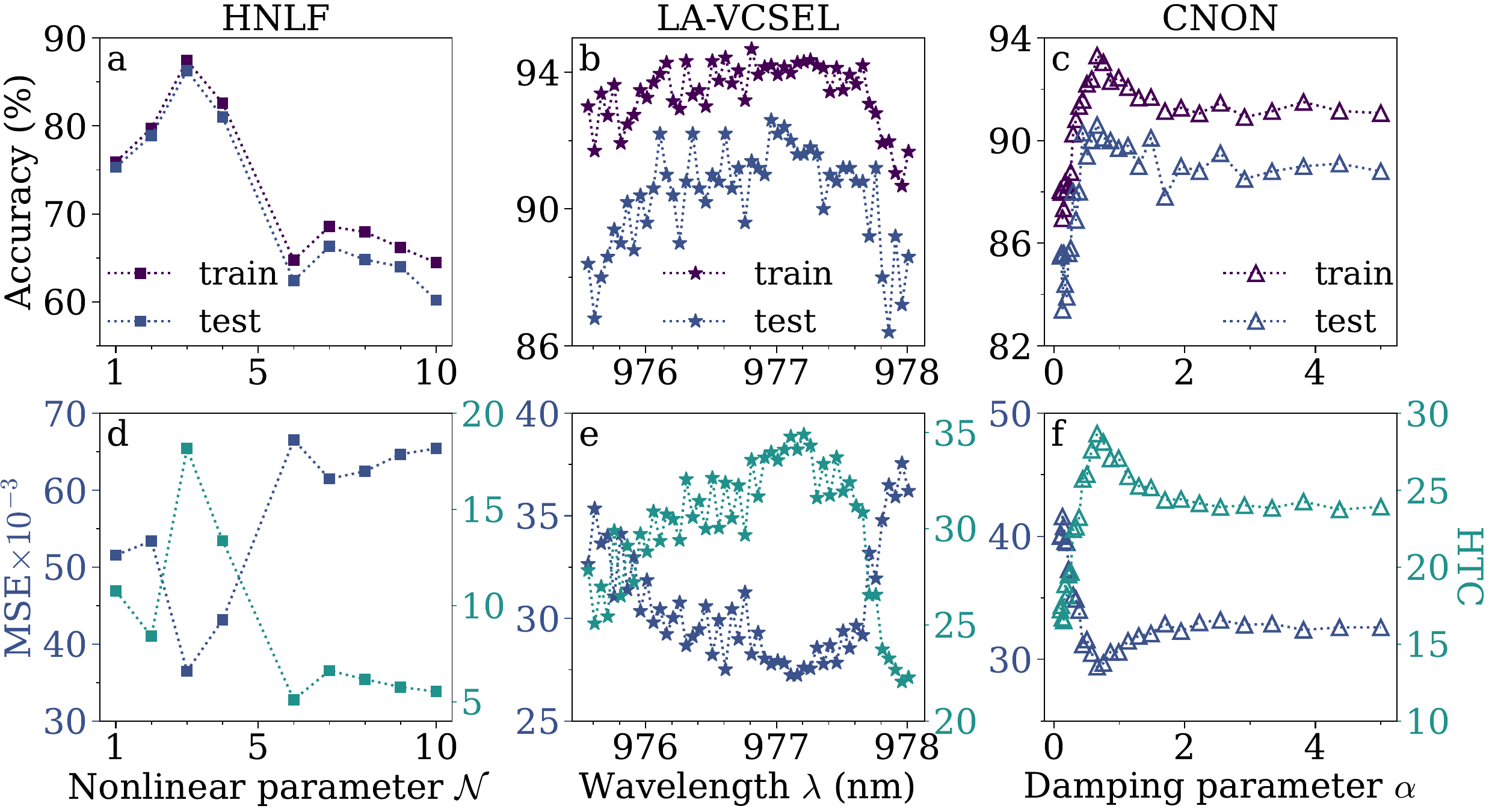}
\caption{\textbf{Hotelling trace criterion as a class separability metric for physical neural networks.} PNN performance represented by test classification accuracy using the MNIST dataset is shown for \textbf{(a)} HNLF experimental data versus the nonlinear parameter $\mathcal{N}$, \textbf{(b)} LA-VCSEL experimental data versus the wavelength $\lambda$, and \textbf{(c)} CNON numerical simulation data versus the damping parameter $\alpha$. \textbf{(d, e, f)} Corresponding dual-axis plots of test MSE (in blue) and $\mathrm{HTC}$ (in green) for HNLF \textbf{(d)}, LA-VCSEL \textbf{(e)}, and CNON \textbf{(f)}. 
}
\label{fig:HTC_PNNs}
\end{center}
\end{figure}
 
For the HNLF-based PNN, classification accuracy is evaluated as a function of the input pulse power, expressed via the system's nonlinear parameter $\mathcal{N}$, see Fig.~\ref{fig:HTC_PNNs}\textbf{(a)}.
 Peak accuracy is obtained at an intermediate value $\mathcal{N} \sim 3$, beyond which performance degrades markedly with increasing input power.
 A qualitatively similar optimum is found in the LA-VCSEL experiment, where the input power is held constant while the detuning between the injection laser and the LA-VCSEL resonance is varied, see Fig.~\ref{fig:HTC_PNNs}\textbf{(b)}.
 Optimal classification accuracy is reached at $\lambda \sim 977.2$~nm, at which the injection laser is resonant with the largest number of transverse LA-VCSEL modes~\cite{Skalli2022}.
 For the CNON, the relevant hyper-parameter is the oscillator damping coefficient $\alpha$ appearing in Eq.~\eqref{eq:PNN_osc_eq}, and a clear performance optimum emerges close to the underdamped regime at $\alpha \sim 0.6$, see Fig.~\ref{fig:HTC_PNNs}\textbf{(c)}.

However, classification accuracy, is a thresholded quantity
in which a PNN's output is binarized into a 1 for the output with the highest amplitude, and a 0 for all others.
 Such binarization is not differentiable and hence it is not a viable loss metric used during training.
 To establish the consistency between task performance and our proposed metric, Figs.~\ref{fig:HTC_PNNs}\textbf{(d, e, f)} show the MSE loss alongside the HTC metric computed for each of the three systems and for shown ranges of their physical parameters.
 Crucially, all results reported here have also been confirmed for using cross entropy loss.
 In contrast to prior studies employing PCA-based dimensionality estimates~\cite{hary2025,Skalli2022}, we find a clear and consistent negative correlation between the MSE loss and the HTC across all three PNNs as a function of their respective physical parameters, indicating that the HTC tracks classification task-relevant computational capacity rather than merely the total available dimensionality.

Figure~\ref{fig:HTC_powerlaw} shows the classification loss as a function of the HTC metric on a double logarithmic scale, revealing a clear power law scaling across both the MNIST digit and fashion MNIST benchmarks.
 For MNIST, the dataset includes experimental measurements from the HNLF and the LA-VCSEL, complemented by numerical simulations of the HNLF and the CNON.
 The fashion MNIST results, being a more demanding benchmark, are currently limited to numerical simulations of the HNLF and CNON systems.
 While the dependence of MSE and HTC on individual physical hyperparameters significantly differs from system to system, see Fig.~\ref{fig:HTC_PNNs}$\mathbf{(a-f)}$, once plotted in the common loss-HTC plane the data collapse onto two distinct power laws, one for each benchmark task (black dashed lines in the inset of Fig.~\ref{fig:HTC_powerlaw}).

 \begin{figure}[H]
\begin{center}
\includegraphics[width=0.9\linewidth]{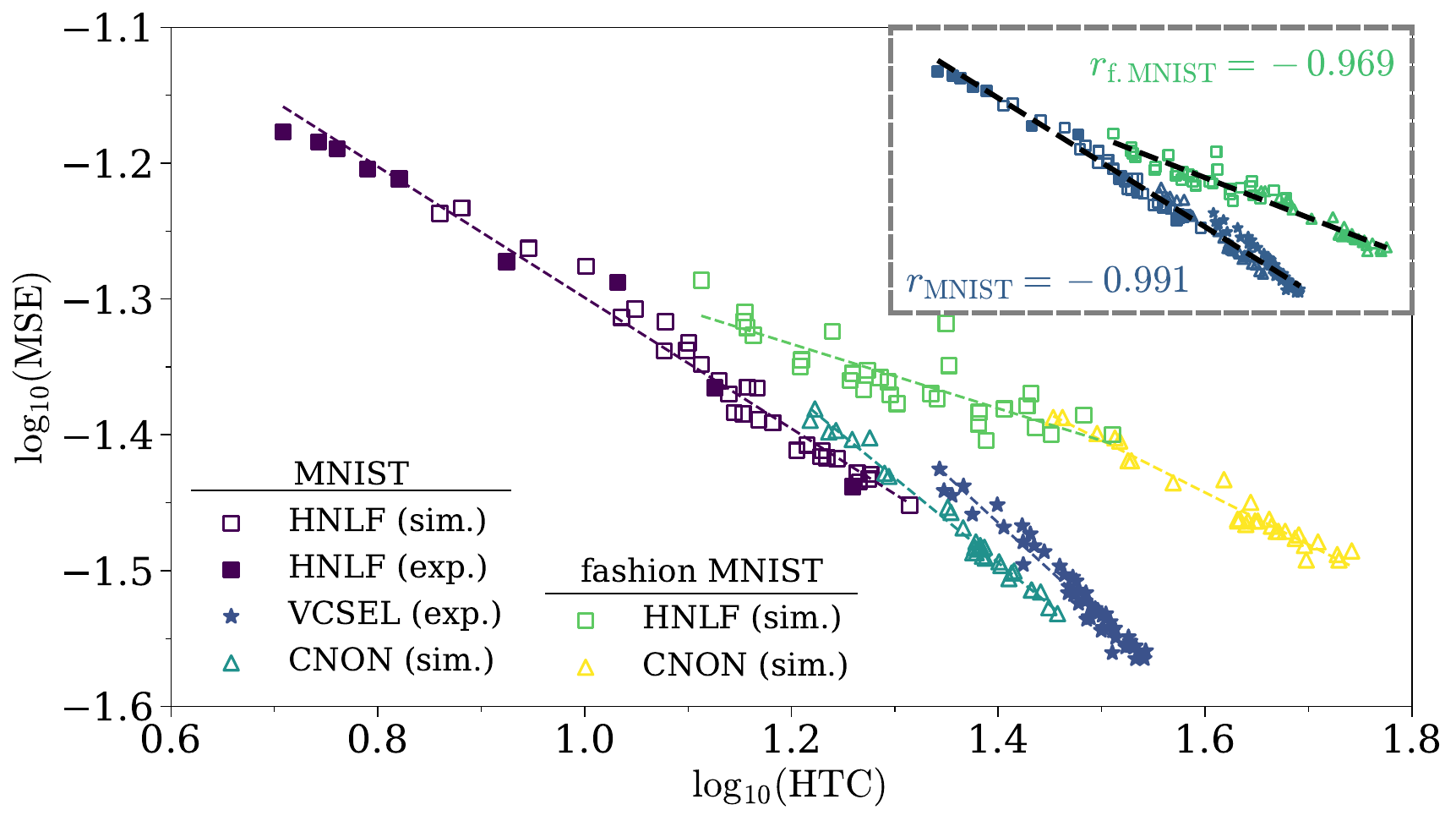}
\caption{\textbf{Observed power law scaling of classification performance with HTC across PNN platforms.} 
Double logarithmic representation of test MSE versus HTC for the three PNN platforms: HNLF, LA-VCSEL, and CNON, evaluated on MNIST and fashion MNIST benchmark datasets. Marker shapes identify individual datasets as indicated in the legend, while filled and open markers distinguish experimental from numerical simulation data, respectively. 
Dashed colored lines show power law fits for each dataset. 
The inset figure in the top right corner shows the overall power laws (black-dashed lines) found for the two tasks (blue and green markers represent MNIST and fashion MNIST datasets, respectively) established with Pearson correlation coefficients of $r_{\mathrm{MNIST}}=-0.991$ and $r_{\mathrm{f. MNIST}}=-0.969$.}
\label{fig:HTC_powerlaw}
\end{center}
\end{figure}

Two features of this scaling are particularly noteworthy.
 First, the power law exponents ($\mathrm{MSE} \propto \mathrm{HTC}^{\zeta}$, with $\zeta_{\textrm{MNIST}}=-0.47$ and $\zeta_{\textrm{f.MNIST}}=-0.29$) differ between the two tasks.
 We interpret this difference as a quantitative measure of their respective computational difficulty, where the easier MNIST classification task results in a steeper power law slope.
 Second, establishing a precise quantitative agreement between HNLF experiments and numerical simulations is challenging and was only partially achieved here.
 Nevertheless, both experimental and simulated data clearly fall onto the same power law, providing strong evidence that the power law's  slope is a property of the task rather than of the specific physical implementation, which is confirmed by the excellent agreement in power law slope for the same task across different PNNs.
 To quantify the quality of task-resolved power law fits, we group all data by benchmark and fit each to a single scaling relation, obtaining Pearson correlation coefficients of $r_{\mathrm{MNIST}} = -0.991$ and $r_{\mathrm{f.\,MNIST}} = -0.969$ for MNIST and fashion MNIST, respectively (inset of Fig.~\ref{fig:HTC_powerlaw}).

\subsection*{Performance analysis of deep PNNs}

The system-agnostic power law established above naturally raises the question of whether HTC can provide equally transparent insight into the internal workings of more complex, multi-layer PNNs and to highlight architectural weaknesses already during training.
 Here we use a three hidden layer CNON PNN with $N = 50$ neurons per layer, which is fully trained using error backpropagation.
 For each training epoch we determine a per-layer HTC value and the entire PNN's MNSE, turning HTC into a dynamic diagnostic of training progress, and Fig.~\ref{fig:HTC_powerlaw_depth}(a) shows the evolution of HTC across the three hidden layers for 20 training epochs after which training has converged.
 All three layers exhibit an initial increase in HTC, reflecting the early reorganisation of internal representations.
 However, the trajectories quickly diverge: the first hidden layer's HTC saturates within a few epochs and subsequently undergoes a steady decline, while the second and third layers' HTC continue to increase for significantly more epochs before stabilising at their respective maximum.
 The saturation levels are also markedly different, with hidden layer~3 reaching an HTC roughly five times larger than that of hidden layer~1 at convergence, hinting at a strong imbalance in their contribution to class separability of the overall PNN.
 
The consequences of this imbalance become explicit when the per-layer HTC trajectories are projected onto the double-logarithmic loss-HTC plane, see Fig.~\ref{fig:HTC_powerlaw_depth}(b).
 Importantly, here the HTC is associated to the entire PNN's loss performance, hence all the layers are linked to the same loss per epoch.
 Hidden layers~2 and~3 adhere excellently to the power law scaling established for the CNON in Fig.~\ref{fig:HTC_powerlaw}, whereas hidden layer~1 strongly diverges from it.
 This deviation we interpret as a fingerprint of an underutilised layer: refinement of its internal representations are not contributing to the overall PNN's performance.
 This is consistent with the well-known diminishing returns in the MNIST task when adding depth to fully connected networks without architectural mitigations such as convolutional layers.
 
The practical implication is equally important: the performance benefit in the MNIST tasks of a three over a two-layer CNON is marginal, with classification accuracy increasing only from $96.3\,\%$ to $96.8\,\%$.
 The HTC analysis provides a clear explanation, showing that the effective computational depth of the network is closer to two than three.
 Beyond this specific case, the result illustrates a broader capability of HTC for PNNs, but also for ANNs in general: by tracking it layer-by-layer throughout training, one obtains a quantitative and interpretable diagnostic of training and architectural efficiency, capable of identifying which layers are well trained, which are underutilised, and whether adding further depth is likely to yield meaningful gains.

\begin{figure}[H]
\begin{center}
\includegraphics[width=0.8\linewidth]{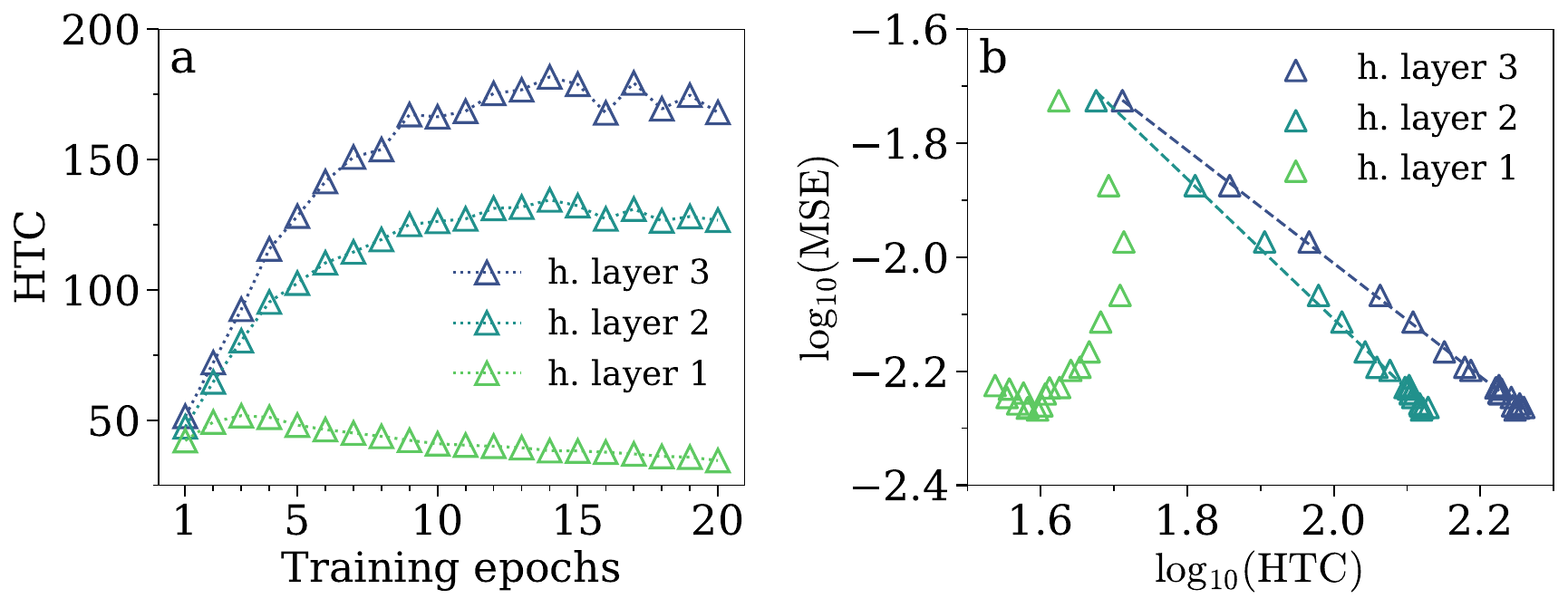}
\caption{\textbf{Layer-resolved evolution of the HTC separability in a fully-trained 3-layer CNON versus the training epochs.} \textbf{(a)} HTC computed from the test hidden-layer matrix $\mathbf{H}_\mathrm{test}$ as a function of training epoch for each of the three hidden layers of a CNON trained on the full MNIST dataset (70,000 images). Despite comparable initial values, the layer HTC metrics diverge with training: the first hidden layer saturates within the first few epochs and subsequently declines, while the second and third hidden layers continue to increase over a substantially larger number of epochs before saturating at significantly higher HTC values. \textbf{(b)} Log-log representation of test MSE versus HTC for each hidden layer, with individual points corresponding to different training epochs. Representations formed in hidden layers~2 and~3 closely align to the universal power law scaling established in Fig.~\ref{fig:HTC_powerlaw} (dashed lines), confirming that these layers are being effectively used during training. By contrast, the first hidden layer deviates strongly from this power law for larger numbers of epochs.}
\label{fig:HTC_powerlaw_depth}
\end{center}
\end{figure}

\section*{Discussion}
\label{sec:discussion}

The results presented in our work establish HTC as a powerful metric for the analysis, comparison, and prediction of PNN performance.
 A practically significant property of HTC is that it does not require training a particular physical PNN realization.
 Rather, it can be computed from their response to injected data only, making it a one-shot or few-shot diagnostic of a physical system's suitability for a given task.
 This stands in stark contrast to the conventional evaluation workflow, in which performance can only be assessed after the full, often costly, training procedure has been completed \cite{momeni2025training}.
 
The power law relationship between HTC and classification loss therefore opens a practically powerful route to performance prediction.
 Once the scaling exponent has been established by training a small number of PNN instances, subsequent performance predictions for systems of the same task can be made from HTC measurements alone.
 The excellent Pearson correlation coefficients obtained for both MNIST ($r_{\mathrm{MNIST}} = -0.991$) and fashion MNIST ($r_{\mathrm{f.\,MNIST}} = -0.969$) confirm that the power law is accurate enough to support quantitative predictions rather than merely qualitative trends.
 This is of particular relevance given the rapidly expanding landscape of PNN platforms, where the cost of exhaustive training-based benchmarking quickly becomes prohibitive~\cite{shastri2021photonics,wright2022deep,momeni2025training}.
 The system-agnostic nature of the HTC power law has direct consequences for how such future PNNs can be designed and evaluated.
 Rather than optimising a physical substrate for a specific task through trial-and-error training, HTC provides a forward-looking figure of merit: a physical system with higher HTC will, to a good approximation, achieve lower classification error on the assigned task, irrespective of the specific nonlinear mechanism or substrate class.
 This reframes the hardware design problem, from maximising task-specific accuracy after training, to maximising task-conditioned separability before training.
  
Beyond PNNs, the HTC framework offers a complementary perspective on ANN training dynamics.
 The layer-wise HTC analysis of the CNON reveals that gradient-based optimisation distributes representational capacity unevenly across depth, with an earlier layer systematically underutilised relative to later ones, a signature that is invisible to standard monitoring leveraging only on the global system's loss.
 This points to a broader role for presentation geometry-based metrics as diagnostic tools during ANN and PNN training rather than relying solely on loss values combined with costly training for an increasing number of layers.
 Tracking the evolution of per-layer HTC provides a quantitative, layer-resolved measure of how efficiently each part of the network is being trained.
 HTC thus offers a principled diagnostic for identifying undertrained layers and, in future work, could inform adaptive training strategies that selectively redirect optimisation effort.
 
The present work establishes one end of a theoretical framework for PNN classification performance.
 A natural and important next step is to close the loop from the other direction by by investigating potential upper HTC bounds and if these are derivable from task-agnostic properties such as previously discussed PCA~\cite{skalli2022computational,hary2025}, computational 
 dimensionality~\cite{dambre2012information} and eigentasks~\cite{tureci2023}.
 This would yield a hierarchy of bounds: task-agnostic methods establish an upper limit, which can then act as guidance for training algorithms to push HTC towards those limits.
 Crucially, these response dimensionalities of a physical system's are itself determined by fundamental physical parameters, which are potentially computable from first principles or measurable without any data-driven optimisation.
 This final step would complete the theoretical loop: from fundamental physical properties of a substrate, via task-agnostic dimensionality bounds towards task-specific performance predictions via the HTC power law.
 Such a framework would transform PNN design from an empirical search into a theoretically grounded discipline, and provide a foundation for establishing universal scaling laws for physical computing  \cite{jaeger2023toward} analogous to those now driving progress in large-scale digital AI~\cite{kaplan2020scaling,hoffmann2022training}.

We note that the present work constitutes a starting point rather than a conclusion.
 To further consolidate our findings, HTC studies should be extended along three directions: (i) to a broader variety of PNN substrates; (ii) to a wider range of benchmark classification datasets; (iii) to physical systems capable of spanning a significantly larger dynamic range within the double-logarithmic scaling law.
 First strides towards regression have been undertaking in PNNs with Kolmogorov-Arnold topology, where state-space packing density was demonstrated to exhibit similar power-law scaling when ranking synapses and architectures in regression tasks~\cite{taglietti2026learning}.
 Beyond scope, an important conceptual limitation also remains: the current formulation of HTC, built on cluster-wise normalised distances, is naturally suited to classification tasks.
 A possibility of its extension to regression problems - or the development of an alternative metric serving the same role - represents an open and pressing challenge for future work.

\newpage

\section*{Methods}

\subsection*{The HNLF-based PNN: Experiment and numerical simulations}
\label{sec:HNLF methods}
For the case of the HNLF, a 125-fs laser 1559.4 nm, 40MHz fiber laser is integrated with a 4-f Fourier domain pulse shaper where a spatial light modulator (SLM) adjusts the spectral phase. The shaped pulses propagate through 5 meters of normal dispersion HNLF and the output spectra are analysed using an optical spectrum analyser (OSA). Input information is encoded by imposing a phase profile on the input spectral amplitude [see Fig.~\ref{fig:PNNs}(b)]. The specific experimental parameters are shown in Table~\ref{tab:parameters}. 

To perform the numerical simulations of the HNLF-based PNN we solve the GNLSE in the form shown in Eq.~(\ref{eq:GNLSE_main}) using standard Fourier split-step method with the nonlinear part treated using Runge-Kutta scheme~\cite{Dudley-2006, Agrawal-2019}. We retain dispersion coefficients up to the fourth order to match the experimental dispersion curve and use the following nonlinear response function $R(t) = (1-f_R)\delta(t)+f_R h_R(t)$, with Raman fraction $f_R = 0.18$, $\delta(t)$ denoting the delta function, and and $h_R$ being the experimental Raman response of fused silica~\cite{Dudley-2006}. To explore the performance of the fibre-based PNN we investigate various dynamical regimes by varying the phase modulation depth $\phi_0$ used for encoding and the nonlinear parameter of the input pulse $\mathcal{N}=\sqrt{\gamma P_0 T_0^2/|\beta_2|}$ that determines the interplay between nonlinear and dispersive effects during propagation. Following Ref.~\cite{hary2025}, we first perform PCA on the raw image dataset and retain a few first principal components, which we encode using equidistant wavelength channels within an input bandwidth.
In addition, we incorporate the semiclassical quantum noise model~\cite{Brainis-2005,Dudley-2006} and include a Raman noise source. To simulate a realistic readout consistent with experimental conditions, we convolve the output spectra with a Gaussian filter and add a noise background (further details of the numerical simulation scheme can be found in~\cite{Ermolaev-2025}). The simulation parameters are selected to approximate the experimental conditions~\cite{hary2025} and are detailed in Table~\ref{tab:parameters}.

\begin{table}[htb!]
\caption{Experimental and numerical simulation parameters for considered HNLF system.}
\label{tab:parameters}
\centering
\begin{tabular}{l c}
\hline
Experimental parameters & Value \\
\hline

Fibre length $L$ & $5~\mathrm{m}$ \\

Pulse duration $T_0$ & $\sim 125~\mathrm{fs}$ \\

Pump wavelength $\lambda_{\mathrm{pump}}$ & $1559.4~\mathrm{nm}$ \\

Spectral range & $1500 - 1850~\mathrm{nm}$ \\

Encoding bandwidth & $\sim 37~\mathrm{nm}$ \\

Number of encoded PCs & $20$ \\

Nonlinear coefficient $\gamma$ & $10.8 \times 10^{-3}~\mathrm{W^{-1}m^{-1}}$ \\

Group-velocity dispersion $\beta_2$ & $1.3 \times 10^{-27}~\mathrm{s^2\,m^{-1}}$ \\

Third-order dispersion $\beta_3$ & $7 \times 10^{-42}~\mathrm{s^3\,m^{-1}}$ \\

Fourth-order dispersion $\beta_4$ & $2.5 \times 10^{-55}~\mathrm{s^4\,m^{-1}}$ \\

\hline
Additional numerical simulation parameters \\
\hline

Nonlinear parameter $\mathcal{N}$ & $ 1-10 $ \\

Phase modulation depth $\phi_0$ & $ \pi/10 - \pi $ \\

Number of grid points & $2048$ \\



Readout convolution window & $0.5~\mathrm{nm}$ \\

Readout noise & $-40~\mathrm{dB}$ \\

\hline
\end{tabular}
\end{table}

\subsection*{The semiconductor laser PNN}

The experimental LA-VCSEL-MNIST dataset was pre-processed as follows. First, a reference mask is generated from the first dataset by averaging 100 randomly selected images and threshold the resulting mean image and applied to all dataset. For each sample, the two-dimensional VCSEL response image was flattened into a one-dimensional feature vector. Pixels outside the mask are removed.
The optical PNN used here is based on the spatial multiplexing of modes in a semiconductor LA-VCSEL laser introduced in \cite{skalli2022computational,porte2021complete}. A digital micromirror device (DMD) encodes the input data through spatial modulation of a single mode injection laser intensity. The resulting optical field is injected into the LA-VCSEL, whose perturbed spatial mode profile under optical injection constitutes the neurons states of the PNN. These PNN activations are then imaged via a camera to allow for offline training.\par
The experimental LA-VCSEL dataset was pre-processed using a fixed spatial masking procedure to ensure consistency across all wavelength measurements. First, a reference mask is generated from by averaging 100 randomly selected images and threshold the resulting mean image. A binary mask is used to represent the active region (inside the illuminated LA-VCSEL) and inactive region (outside the LA-VCSEL). Each sample is then flattened into a one-dimensional vector, where pixel outside the reference mask were removed, retaining only the informative region. 

\subsection*{Coupled nonlinear oscillator network PNN}
\label{sec:CNON}

The generic oscillator network with homogeneous activation functions used in our analysis is  described by the following model:

\begin{equation}\label{eq:PNN_osc_eq}
\dfrac{d\mathbf{h}^{[l]}(t)}{dt} = 
-\alpha\, \mathbf{h}^{[l]}(t)
+ \sin\!\left(
\mathbf{W}_{\mathrm{rec}}^{[l]} \mathbf{h}^{[l]}(t)
+\mathbf{W}_{\mathrm{in}}^{[l]} \mathbf{h}^{[l-1]}(t)
+ \mathbf{b}^{[l]}
\right), \qquad l>1
\end{equation}
\begin{equation}\label{eq:PNN_osc_out}
\mathbf{y}(t)
=
\mathbf{W}_{\mathrm{out}}\,\mathbf{h}^{[L]}(t)
+\mathbf{b}_{\mathrm{out}}
\end{equation}

The state of the hidden layer $l$ is described by the vector $\mathbf{h}^{[l]}$, which evolves continuously according to Eq.~(\ref{eq:PNN_osc_eq}). Unlike conventional discrete-time recurrent neural networks, the network dynamics are integrated numerically using Euler integration with a sufficiently small time step. The physical realisation of a single layer is shown in Fig.~\ref{fig:HTC_PNNs}\textbf{(d)}.

To satisfy physical constraints, the recurrent coupling matrix $\mathbf{W}{\mathrm{rec}}$ is symmetric, fixed, $80\%$ sparse, and normalized by its largest eigenvalue. The parameter $\alpha$ controls the damping strength. For memory-independent tasks, each input sample is held constant until the system converges to a fixed point. The fixed-point state of the final layer is then mapped to the network output according to Eq.~(\ref{eq:PNN_osc_out}). Here, $\mathbf{W}{\mathrm{in}}^{[l]}$ and
$\mathbf{W}{\mathrm{out}}$ denote the input and output weight matrices, respectively.

\newpage

\subsection*{Data analysis and PNN training}
\label{sec:analysis and training}

To assess the performance of different PNN considered in this work, we use the MNIST and fashion MNIST benchmark classification tasks. In the case of HNLF-based PNN, we follow the procedure described in Ref.~\cite{hary2025}. Specifically, we first perform the PCA using the dataset containing $M$ images and retain 20 first principal components corresponding to each image, which are next used for encoding. For LA-VCSEL and CNON PNNs we use raw 28 by 28 pixel images from both datasets as the input data, respectively. In all the cases, the systems output forms hidden layer matrix $\mathbf{H} \in \mathbb{R}^{M \times N}$ that we split into training, validation and test datasets denoted $\mathbf{H}_{\mathrm{train}}\in \mathbb{R}^{m_\mathrm{train} \times N}$, $\mathbf{H}_{\mathrm{val}}\in \mathbb{R}^{m_\mathrm{val} \times N}$, and $\mathbf{H}_{\mathrm{test}}\in \mathbb{R}^{m_\mathrm{test} \times N}$, respectively (here $M=m_\mathrm{train}+m_\mathrm{val}+m_\mathrm{test}$).

During the ELM training, the output weight matrix $\mathbf{W}_{\mathrm{out}} \in \mathbb{R}^{N \times K}$ (with $K=10$ in our case) can be trained using the ridge regression 
\begin{equation}
    \tilde{\mathbf{W}}_{\mathrm{out}} = (\mathbf{H}_{\mathrm{train}}^{T}\mathbf{H}_{\mathrm{train}} + \lambda_R \mathbf{I})^{-1}\mathbf{H}_{\mathrm{train}}^{T} \mathbf{Y}_{\mathrm{train}}, 
\end{equation}
where $\lambda_R$ is the ridge regularisation parameter and $\mathbf{Y}_{\mathrm{train}} \in \mathbb{R}^{m_\mathrm{train} \times K}$ is the one-hot encoded training target matrix. The optimal value of $\lambda_R$ is selected via a grid search over the validation set. Next, the classification accuracy is evaluated by applying $\tilde{\mathbf{W}}_{\mathrm{out}}$ to the to the test hidden-layer representations $\mathbf{H}_\mathrm{test}$. The test mean squared error (MSE) is computed as follows: 
\begin{equation}
    \mathrm{MSE}_\mathrm{test} = \dfrac{1}{K \,m_{\mathrm{test}}} \left\|\mathbf{Y}_\mathrm{test} - \mathbf{H}_\mathrm{test}\tilde{\mathbf{W}}_{\mathrm{out}}\right\|_{F}^{2},
\end{equation}
where $\mathbf{Y}_{\mathrm{test}}\in \mathbb{R}^{m_\mathrm{test} \times K}$ is the test target matrix and $||\cdot||_{F}$ denotes the Frobenius norm. The test classification accuracy can be subsequently estimated as the fraction of correctly predicted labels by comparing $\mathrm{argmax}_{k}(\mathbf{Y}_\mathrm{test})$ and $\mathrm{argmax}_{k}(\mathbf{Y}_\mathrm{pred})=\mathrm{argmax}_{k}(\mathbf{H}_\mathrm{test} \tilde{\mathbf{W}}_{\mathrm{out}})$, where $\mathrm{argmax}$ function is applied with respect to the classes. 

The considered PNNs have noticeably different structure of the output state vector $\mathbf{h}_{i}$ due to the experimental constraints or details of numerical simulations. Before performing the analysis we first linearly interpolate the data representing the hidden-layer matrix $\mathbf{H}$ to a uniform dimensionality of $N = 600$ for all the cases. This is done in order to maintain the consistency in evaluation of the MSE, classification accuracies, as well as, the HTC metric. In principle, data can be linearly interpolated to any number of neurons $N$, ensuring that the hidden-layer representation retains sufficient dimensionality.  

The details of the HTC calculation are provided in Sec.~\ref{sec:HTC}. We note that in the case of ELM training and analysis we use the entire hidden-layer matrix $\mathbf{H}$ to compute HTC due to the limited number of samples. To ensure numerical stability, the within-calss scatter matrix is usually regularised as $\mathbf{S}_W+\epsilon\mathbf{I}$, where $\epsilon \ll 1$ and $\mathbf{I}$ is the identity matrix. The described data pre-processing, analysis and training are applied to all the data shown in Figs.~\ref{fig:HTC_PNNs}-\ref{fig:HTC_powerlaw}. For CNON simulation data, as well as for the HNLF experimental and numerical simulation data we use a total of 10000 MNIST images split into 8000, 1000, and 1000 training, validation and test samples, respectively. The LA-VCSEL experimental dataset consists of 5000 MNIST images that we split into 4000, 500, and 500, respectively.

For the fully-trained single- and multi-layer CNON results, we use the complete MNIST dataset of 70,000 images (60,000 training and 10,000 test samples) with a batch size of 512 images. Unlike in the ELM case discussed above, the fully-trained multi-layer CNON exhibits an increased sensitivity to overfitting. To account for this, the HTC metric is computed using the test hidden-layer matrix $\mathbf{H}_\mathrm{test}$ throughout this analysis (see Fig.~\ref{fig:HTC_powerlaw_depth}).

\newpage

\section*{Acknowledgments}

We acknowledge funding from the European Union’s Horizon 2020 research and innovation programme with European Research Council (ERC) (Consolidator Grant, grant agreement No. 101044777 (INSPIRE)), with the Marie Skłodowska-Curie grant agreement No. 860830  (POST DIGITAL) and No. 101169118 (POSTDIGITAL+) by the French Investissements d’Avenir program, the French Agence Nationale de la Recherche (ANR) (Contract No. ANR-17-EURE-0002, ANR-15-IDEX-0003, ANR-17-EURE-0002, ANR-21-CE47-0021) as well as support by the Region Bourgogne Franche Comte, the Academy of Finland (318082, 320165, 333949), by the German Research Foundation (via Phase III of the Collaborative Research Center project 787), as well as JSPS KAKENHI (Grant No. JP22H05198) and JST CREST (Grant No. JPMJCR24R2).

\section*{Data and code availability}

The data and code used to generate and analyze the MNIST classification results presented in this work are publicly available at \url{https://github.com/AndreiErmolaev/ELM-classifier-utils}. The code used for simulations of the coupled nonlinear oscillator network (CNON) is publicly available at \url{https://github.com/ASkalli/nonlinear_oscillator_network}. All other data and numerical simulation codes that support the findings of this study are available from the corresponding author upon reasonable request.

\section*{Competing interests}
Competing interests: We declare that none of the authors have competing financial or non-financial interests as defined by Nature Portfolio.

\section*{Author contributions}

The concept of leveraging state-space embedding, its dimensionality covered linked to input data as well as its class overlapped was developed by AVE, MH and DB.
 The link from these concepts to the HTC was established by AVE, and developed for the application to PNNs with input from MH under supervision of DB.
 MH carried out the PNN-HNLF experiment under the supervision of DB, JD and GG. 
 AS carried out the LA-VCSEL experiment and developed the CNON framework under the supervision of DB.
 MG fabricated and characterized the LA-VCSELs devices under the supervision of JL, SR provided the fabrication infrastructure.
 AVE and JMD developed the HNLF-based PNN simulation pipeline, and AVE performed the numerical simulations. 
 AVE performed the HTC analysis of the HNLF-based PNN and CNON, and MH performed the HTC analysis of the LA-VCSEL experimental data.
 DB, AVE and MH wrote the manuscript with contributions from all authors.

\bibliography{PNN_bib}

\end{document}